\documentclass[reprint,amsmath,amssymb,aps,floatfix,
]{revtex4-2}
\usepackage{graphicx}% Include figure files
\usepackage{dcolumn,xcolor}% Align table columns on decimal point
\usepackage{bm}% bold math
\begin{document}

\preprint{APS/123-QED}

\title{Hyperelastic swelling of stiff hydrogels}% Force line breaks with \\
%\thanks{A footnote to the article title}%

\author{Jing Wang}
\author{Justin C. Burton}%
 \email{justin.c.burton@emory.edu}
\affiliation{Department of Physics, Emory University.}

\date{\today}% It is always \today, today,
             %  but any date may be explicitly specified

\begin{abstract}
Hydrogels are biphasic, swollen polymer networks where elastic deformation is coupled to nanoscale fluid flow. As a consequence, hydrogels can withstand large strains and exhibit nonlinear, hyperelastic properties. For low-modulus hydrogel and semiflexible biopolymer networks, previous studies have shown that these materials universally contract when sheared on timescales much longer than the poroelastic relaxation timescale. Using rheological and tribological measurements, we find that stiff polyacrylamide and polyacrylic acid hydrogels, with moduli of order $\sim$ 10-100 kPa, exclusively swell (dilate) when sheared. The poroelastic relaxation process was examined using strain-controlled compression, indicating a volumetric diffusion constant of order 10$^{-9}$ m$^2$/s. Upon shearing, we observed an increase in normal stress that varied quadratically with shear strain, and persisted for hours. Moreover, we show that this dilatant behavior manifests as swelling during tribological sliding, imbibing the hydrogel with fluid. We suggest that this inherent, hyperelastic dilatancy is an important feature in all stiff hydrogels, and may explain rehydration and mechanical rejuvenation in biological tissues such as cartilage. 
\end{abstract}

%\keywords{Suggested keywords}%Use showkeys class option if keyword
                              %display desired
\maketitle

%\tableofcontents

\section{Introduction}
Hydrogels exhibit complex, time-dependent mechanical properties due to their biphasic nature. Material deformation inherently requires the solvent to flow relative the underlying cross-linked polymer matrix. In recent years, hydrogels have been engineered and investigated for applications such as urban farming \cite{palanivelu2022hydrogel}, cosmetic care \cite{mitura2020biopolymers, sachdev2010facial}, tissue engineering \cite{lee2001hydrogels, silverberg2014structure}, and drug delivery \cite{hamidi2008hydrogel, li2016designing, dreiss2020hydrogel}. In particular, stiff hydrogels with a Young's modulus $>$ 10 kPa exhibit ultra-low friction and have been used as a model system for articular cartilage \cite{sardinha2013tribological, parkes2015tribology, blum2013low, bavaresco2008study, freeman2000friction, kim2016soft}. Recent tribological experiments show an increase of cartilage (hydrogel) volume upon sliding at a smooth interface, a process attributed to ``tribological rehydration'' \cite{burris2017cartilage, burris2019sliding, farnham2020effects, putignano2021cartilage, de2020compliant, kupratis2021comparative}. The rehydration process is velocity-dependent and is claimed to only occur at the hydrogel contact interface as a result of elastohydrodynamic lubrication pressure \cite{moore2017tribological}. In this sense, the application of shear (through mechanical motion, i.e., joint activity) will imbibe the cartilage with fluid, softening and rejuvenating the tissue.

Alternatively, here we will show that rejuvenation and swelling of stiff hydrogels can be ascribed to their bulk, hyperelastic mechanical properties. Specifically, the Poynting effect, which describes the development of stress perpendicular to the direction of shear in a confined elastic material  \cite{poynting1905xxxix, poynting1909pressure, billington1986poynting}. The Poynting effect has been observed in materials such as rubber \cite{vitral2023stretch}, metal wires \cite{poynting1905xxxix}, soft fiber materials \cite{horgan2017poynting, horgan2021effect}, soft polymer gels \cite{shivers2019normal, niroumandi2021finite}, and emulsions \cite{fall2022tuneable}. The change in normal stress, $\Delta\sigma_N$, in hyperelastic materials is captured by the Mooney-Rivlin model \cite{mooney1940theory, rivlin1948large}, which predicts that when sheared at a fixed volume, $\Delta\sigma_N\propto\gamma^2$, where $\gamma$ is the shear strain. The effect can be positive (dilatant) or negative (contractile). In 2007, Janmey \textit{et al.} \cite{janmey2007negative} first reported contractile behavior in semiflexible biopolymer gels such as fibrin, which has been confirmed by subsequent studies \cite{kang2009nonlinear, vahabi2018normal, conti2009cross, licup2015stress}. However, Janmey \textit{et al.} \cite{janmey2007negative} also reported a dilatant behavior in low-modulus, weakly-cross-linked polyacrylamide (PAAm) hydrogels.  While this was attributed to the flexibility of the polymers, instead de Cagny \textit{et al.} \cite{de2016porosity} revealed that poroelastic flow on long timescales can alter the sign of the normal stress response, so that an initial dilatant behavior can transition to a contractile one in PAAm with moduli of order $\sim$ 100 Pa. They further suggested that any confined hydrogel would eventually exhibit negative normal stress in response to shear after sufficient waiting time. However, here we show stiff hydrogels with moduli of order 10-100 kPa \emph{always} exhibit dilatant, hyperelastic behavior, even long after reaching poroelastic equilibrium.

Both contractile and dilatant responses in hydrogels can be predicted based on the change in shear modulus under compression. Following Tighe \textit{et al.} \cite{tighe2014shear,baumgarten2018normal}, when sheared under constant volume, the normal stress response of an elastic solid can be expressed as 
\begin{align}
    \Delta \sigma_N = \frac{1}{2}R_v \gamma^2 + \mathcal{O}(\gamma^4), \label{normal stress tighe}
\end{align}
where $R_v$ is the Reynolds coefficient at a fixed volume: 
\begin{align}
    R_v = E \left( \frac{\partial G}{\partial \sigma_N}   \right)_\gamma   - G. \label{reynolds coeff. tighe}
\end{align}
Here, $G$ is the shear modulus, $E$ is the Young's modulus, and the subscript $\gamma$ implies that the derivative is taken at constant, infinitesimal shear strain. Generally, $G < E$, and the material will exhibit dilatant behavior if the differential in the first term is large enough and positive. For granular packings, $\partial G/\partial\sigma_N>0$, as evidenced by studies of dilatancy in granular materials dating back to 1886 \cite{reynolds1886experiments}, and more recently in a broader class of granular materials such as emulsions \cite{fall2022tuneable}. For spring networks, often used as a minimal model for cross-linked gels, $\partial G/\partial\sigma_N<0$, implying that spring networks universally contract upon compression \cite{baumgarten2018normal}. As mentioned, contractile behavior is indeed observed in semi-flexible biopolymer networks and weakly cross-linked hydrogels \cite{janmey2007negative,de2016porosity}.

We have conducted a series of both surface tribological sliding and bulk rheologial shearing experiments using stiff PAAm and polyacrylic acid (PAA) hydrogels. These hydrogels exclusively dilate in all experiments, in contrast to low modulus gels and the expected behavior of network materials. We first characterized the volumetric diffusion of the hydrogel network upon compression in order to isolate transient relaxation behavior from rheological response. We then illustrate how dilatant behavior is manifested in confined and unconfined tribological sliding experiments. Finally, we used oscillatory shear and strain sweep to directly measure $R_v$, and show that Eq.~\ref{reynolds coeff. tighe}, which is derived from a purely elastic solid, is a reasonable estimate to the dilatant behavior. These results provide a natural explanation for the volumetric expansion of cartilage during sliding, and are applicable to a broader class of stiff synthetic hydrogels and biological tissues.

%Finally, we used high-amplitude oscillatory shear to illustrate how the dilatant behavior of stiff hydrogels deviates from simple, quadratic models of hyperelasticity (Eq.~\ref{normal stress tighe}) over long times due to irreversible, plastic deformation.   

\section{Diffusion-Driven Relaxation }

The hyperelastic response of hydrogels can change in both magnitude and sign during poroelastic relaxation \cite{de2016porosity}. This process is driven by the relative diffusion of the solvent and polymer network, as described by Doi \textit{et al.} \cite{doi2009gel,doi1978dynamics}. The diffusion constant can be written as
\begin{align}
    D = \frac{Ek(1-\nu)}{\mu (1+\nu)(1-2\nu)},
    \label{diffeq}
\end{align}
where $E$ is the Young's modulus of the hydrogel, $k$ is the permeability, $\mu$ is the viscosity of the solvent (water), and $\nu$ is Poisson's ratio \cite{louf2021poroelastic}. We note that $D$ does not represent mass diffusion, but rather diffusion of volumetric strain, $\nabla\cdot\vec{\bf u}$, where $\vec{\bf u}$ is the displacement field \cite{doi2009gel}. %\cite{hong2008theory}. 
For the PAAm and PAA hydrogel spheres used in our experiments, $E\approx35.7$ kPa (Fig.~S1) \cite{supp}, $\nu\approx0.40$, and $k\propto\xi^2$, where $\xi\approx 5-15$ nm is the mesh size of the polymer network \cite{cuccia2020pore}. Thus we expect $D\approx2-20\times10^{-9}$ m$^2$/s. A full description of the hydrogel preparation methods can be found in the Supplemental Materials (SM) \cite{supp}.

\begin{figure}[h]
\centering
\includegraphics[width=0.85\columnwidth]{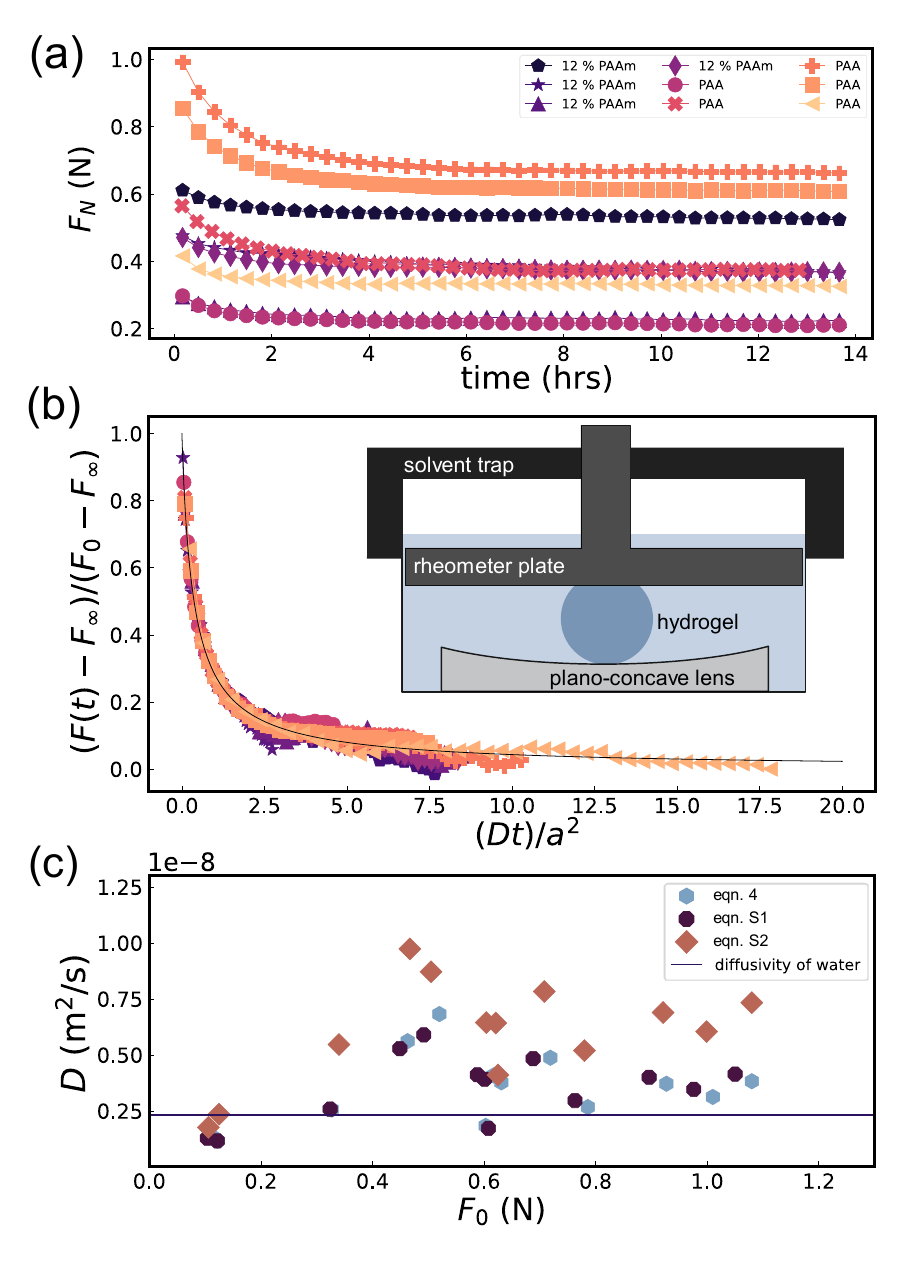}
%{paper-compression 100223.pdf}
\caption{Stiff hydrogels display poroelastic relaxation upon compression. (a) Under fixed strain, the applied normal force relaxed to a constant value after many hours. (b) Assuming a diffusive relaxation process (Eq.~\ref{fitform1}, SM Eq. 1 and 2), the data can be collapsed by scaling both axes. Inset: experimental schematic showing a hydrogel sphere (diameter $\approx$ 1.5 cm) compressed in a rheometer and left to relax under constant normal strain. The concave lens ensured centering of the gel. (c) The extracted diffusion constant was independent of the exact fitting form, and comparable to, or larger than, the self-diffusivity of water (solid line).}
\label{fig:compression}
\end{figure}

We characterized the diffusion process in our hydrogels by examining the relaxation of the normal force under compression using a rheometer (AR 2000, TA Instruments). With a 60-mm parallel plate geometry, hydrogel spheres were compressed on top of a plano-concave lens fully immersed in a water bath (Fig.~\ref{fig:compression}(b) inset). A custom solvent trap was placed on top of the water bath to prevent evaporation over day-long timescales. Each experiment began by applying a constant normal force, $F_0$, which determined the amount of compressive normal strain ($\approx 1-10\%$). We subsequently recorded the normal force response, $F_N(t)$, during relaxation for 14 hours. The normal strain (determined by the rheometer gap size) remained constant during the experiment. Figure \ref{fig:compression}(a) shows $F_N(t)$ for commercial PAA and 12$\%$ PAAm spheres with different levels of initial compression. In all cases, $F_N(t)$ decreased rapidly at early times but eventually reached a plateau at late times. This behavior agrees with numerous previous studies using indentation methods such as the atomic force microscopy to study hydrogel poroelasticity \cite{berry2020poroelastic, hu2010using, kalcioglu2012macro, cuenot2022mechanical, esteki2020new, burris2017cartilage}. 

The data in Fig.~\ref{fig:compression}(a) can be collapsed onto a single curve by scaling the vertical and horizontal axes. The initial and final values of the force determine the vertical axis scaling, whereas time is scaled by $\tau=a^2/D$, where $D$ should be the diffusion constant (Eq.~\ref{diffeq}). The contact area $a$ was measured using particle exclusion microscopy as described in Ref.~\cite{cuccia2020pore}. The diffusion constant can be found by fitting the data to a functional form that can capture the relaxation dynamics \cite{berry2020poroelastic,hu2010using}. To collapse the data in Fig.~\ref{fig:compression}(a), we used the following function, taken from Berry \textit{et al.} \cite{berry2020poroelastic}:
\begin{align}
    \frac{F_N(t) - F_\infty}{F_0 - F_\infty} &= 1 - \frac{2.56(Dt/a^2)^{0.94}}{1+2.56(Dt/a^2)^{0.94}},
    \label{fitform1}
\end{align}
where $F_\infty=F_N(t=\infty)$. This function is empirical, and we found that other fitting functions provide similar results, as described in the SM \cite{supp}. Since $a$ is measured by microscopy, the fitting parameters were $D$, $F_0$, and $F_\infty$. The collapsed data is shown in Fig.~\ref{fig:compression}(b). The extracted diffusion constants ($D$) using Eq.~\ref{fitform1}, Eq.~S1, and Eq.~S2 are shown in Fig.~\ref{fig:compression}(c) as a function of the initial compressive force, $F_0$. The values of $D$ ranged from $1.2-9.7 \times 10^{-9}$ m$^2$/s, which is consistent with the estimate from Eq.~\ref{diffeq}. We note that our measurements are 10-100 times larger than values of $D$ reported from micro-indentation measurements on hydrogel surfaces \cite{berry2020poroelastic,esteki2020new}. Despite this discrepancy, we also note excellent agreement with values of $D$ extracted from swelling experiments of similar-sized hydrogel particles \cite{louf2021poroelastic}.

\section{Tribological Swelling}

Under constant normal load, stiff hydrogels are free to swell in response to an imposed shear stress, for example, frictional sliding against a smooth surface. The shear stress applied at the sliding interface is transmitted into the bulk, resulting in hyperelastic swelling. In confined environments where swelling is inhibited, this effect manifests as an increase in normal stress. In fully-relaxed, compressed hydrogel spheres, we observed sliding-induced swelling using stress-controlled tribology experiments and strain-controlled rheology experiments.

%\begin{figure}[!]
%\centering
%\includegraphics[width=0.95\linewidth]{tribology.pdf}
%\caption{Stiff hydrogels swell during tribological sliding. Plotted data represents vertical displacement vs. time during the initial compression ($t<$ 10 hrs), during sliding at 1.0 cm/s (10 hrs $<t<$ 13 hrs), and post sliding ($t>$ 13 hrs). The contact radius of the PAA sphere was $\approx$ 3 mm under a normal load of 0.2 N. The inset shows an illustration of the stress-controlled tribology setup for displacement imaging. The relaxation and swelling of the hydrogel was monitored by tracking the position of the red circle on the apparatus. A thin layer of water was used to ensure hydration of hydrogel spheres.}

%\label{fig:dottracking}
%\end{figure}

\begin{figure}[h]
\centering
\includegraphics[width=0.95\linewidth]{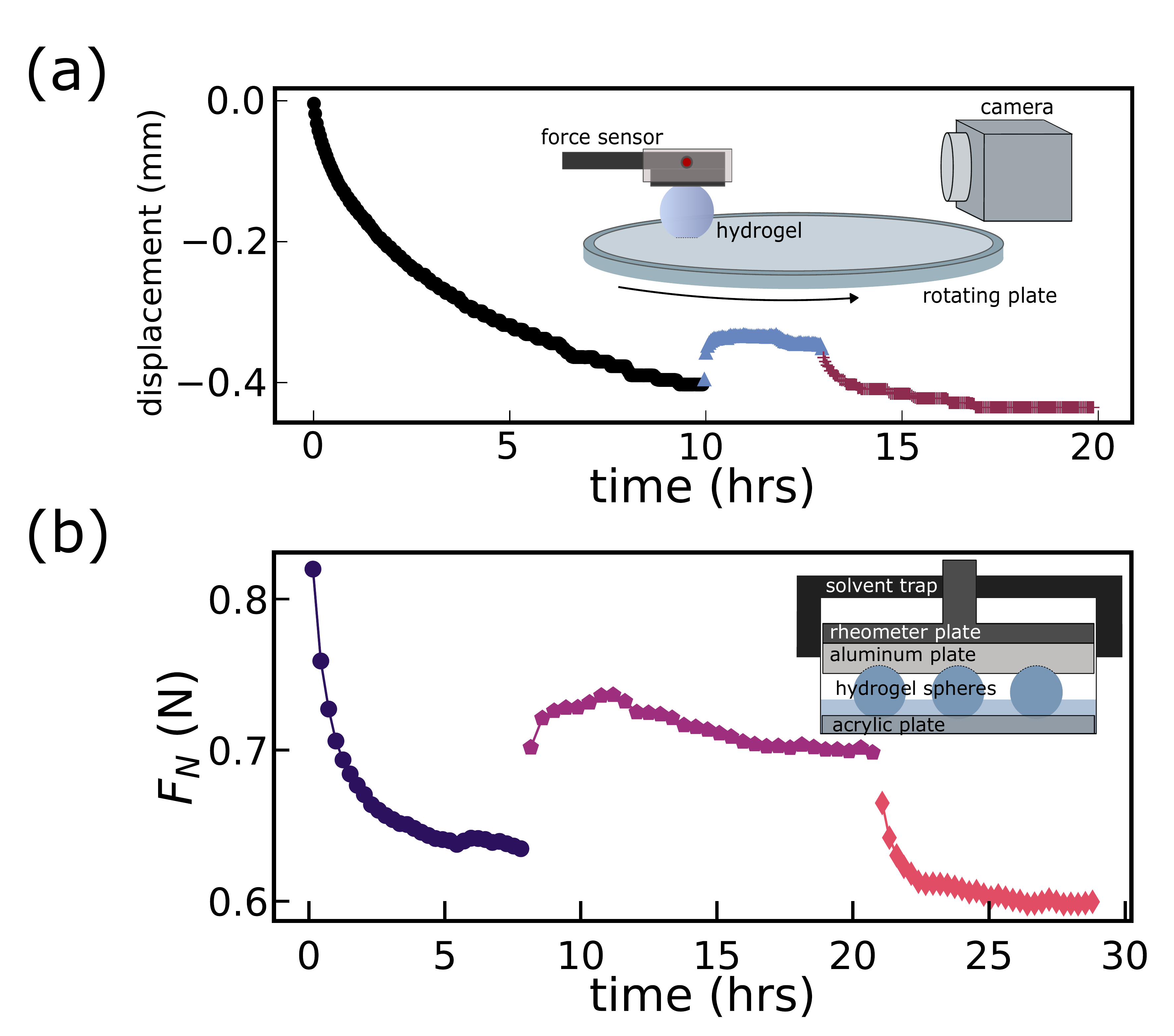}
%{paper-rheotribology 100223.pdf}
\caption{(a) Stiff hydrogels swell during tribological sliding. Plotted data represents vertical displacement vs. time during the initial compression ($t < 10$ hrs), during sliding at 1.0 cm/s (10 hrs $< t < 13$ hrs), and post sliding (t $ > $ 13 hrs). The contact radius of the PAA sphere was $\approx$ 3 mm under a normal load of 0.2 N. The inset shows an illustration of the stress-controlled tribology setup for displacement imaging. The relaxation and swelling of the hydrogel was monitored by tracking the position of the red circle on the apparatus. A thin layer of water was used to ensure hydration of hydrogel spheres. (b) Stiff hydrogels exhibit a positive normal force upon sliding under confinement. The PAA spheres underwent an 8-hour initial relaxation, a period of sliding at a constant sliding velocity (0.3 cm/s), and further relaxation post-sliding. The inset shows the experimental setup, as described in the main text. Hydrogels were surrounded by a thin layer of water to ensure hydration.}
\label{fig:tribology}
\end{figure}

Figure \ref{fig:tribology}(a) inset shows our custom-built tribometer \cite{cuccia2020pore,wu2021relaxation}. Hydrogel spheres were compressed onto an acrylic surface with a constant normal load of $F_N =$ 0.2 N. A spot was marked above the hydrogel sphere and its vertical position was recorded at 1 fps and tracked using image analysis. The time evolution of the vertical displacement is depicted in Fig. \ref{fig:tribology}(a). We define $t = 0$ hrs when the normal load was first applied to the hydrogel sphere. For $0 < t < 10$ hrs, the displacement relaxed to equilibrium, similar to the normal force in Fig. \ref{fig:compression}(a). At $t=10$ hrs, a constant sliding velocity of 1 cm/s was applied for 3 hrs. The vertical displacement increased rapidly by 70 $\mu$m and ultimately reached a plateau. Importantly, upon cessation of sliding at $t=13$ hrs, the vertical displacement continued to relax from the plateau with a similar timescale as the initial relaxation. This demonstrates that swelling occurred over the bulk of the material and not simply at the interface.

In confined environments, the dilatancy of stiff hydrogels under shear or sliding will lead to an increase normal force. To show this, we performed sliding experiments using a strain-controlled rheometer with PAA commercial hydrogel spheres, as depicted in Fig. \ref{fig:tribology}(b) inset. An aluminum plate (diameter = 60 mm) with 3 divots was glued to the top parallel plate of the rheometer. The divots were placed at the vertices of an equilateral triangle, each 20 mm from the plate center. A small amount of modeling clay was used in the divots to prevent undesired rolling of the spheres. Sliding occurred at a smooth acrylic surface at the bottom of each sphere. An initial normal force of $F_0 =$ 0.9 N was applied to compress the hydrogel spheres for 8 hours. Subsequently, we applied a constant sliding velocity $v = 0.3$ cm/s to hydrogel spheres for $\sim$ 13 hrs. Hydrogel spheres were monitored for 8 hours post sliding. The gap size was constant throughout the experiment, corresponding to a normal strain of 3\%. Upon initial compression, the normal force relaxed similarly to Fig.~\ref{fig:compression}(a). During sliding ($8 < t < 21$ hrs), $F_N$ initially increased by nearly 20\% and eventually stabilized around the 16th hour. The post-sliding relaxation observed in Fig.~\ref{fig:tribology}(b) occurred over a similar timescale as the relaxation from the initial compression, providing further evidence of bulk, hyperelastic swelling during sliding.

\section{Hyperelastic dilation from simple shear}

\begin{figure}[!]
\centering
\includegraphics[width=.95\linewidth]{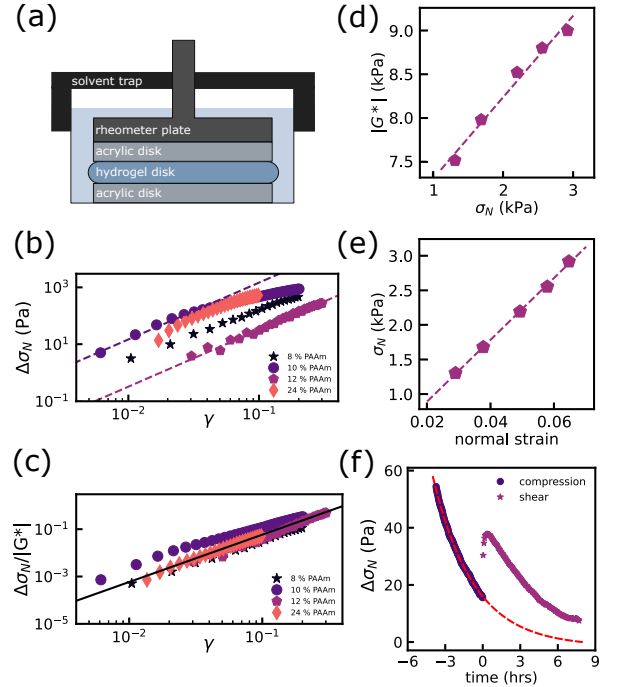}
%{paper-cartoon_hyperelasticity.pdf}
\caption{Oscillatory rheology of hydrogels reveals a hyperelastic dilatancy. (a) Illustration showing a compressed PAAm disk of diameter 60 mm and thickness 6 mm between two acrylic disks affixed with sandpaper. Each hydrogel disk was compressed with an initial normal stress and allowed to relax prior to testing. (b) The change in normal stress ($\Delta \sigma_N$) vs. strain ($\gamma$) for different hydrogel samples during an oscillatory strain sweep at 1 Hz. Each point is an average of 25 s spent at the corresponding strain. Dashed lines represent fitting results using the function $\Delta \sigma_N = R_v\gamma ^2/4$ (Eq.~S8 \cite{supp}). (c) By normalizing $\Delta \sigma_N$ using $|G^*|$ at each value of the strain (see Fig.~S2), the data in panel (b) can be reasonably collapsed. The solid line represents $\Delta\sigma_N/|G^*|\approx 5.9\gamma^2$. (d) Complex modulus $|G^*|$ versus $\sigma_N$ measured via oscillatory shear (1 Hz) at constant shear strain ($\gamma = 0.05$). (e) Normal stress versus normal strain. Young's modulus ($E$) can be calculated from the data (Eqs.~S5-S6 \cite{supp}). For each point in (d-e), the hydrogel was relaxed to equilibrium 3 hrs before testing, and the dashed lines represent linear fits.  (f) $\Delta\sigma_N$ vs. time for an 8\% PAAm disk that was initially compressed to 2000 Pa. A subsequent period of shear stress ($\sigma_s$ = 150 Pa) induced an increase in normal stress during the relaxation. The dashed line is an exponential fit to the relaxation period to estimate $F_\infty$, which is used to determine $\Delta\sigma_N=0$. During shear, the strain remained constant at $\approx$ 20\%, indicating minimal slip at the hydrogel/sandpaper interfaces.}
%{\color{red}(e) Continuous straining at $\gamma = 10\%$ for $\sim$ 10 hours at 1 Hz. An immediately increase of $\Delta \sigma_N$ was observed upon shear and showed no sign of relaxation.} }
\label{fig:normal force vs strain}
\end{figure}

To further show that the dilatant effects observed during sliding are a bulk property, we used oscillatory rheology to quantitatively investigate the strain dependence predicted in Eqs.~\ref{normal stress tighe} and \ref{reynolds coeff. tighe}. We fabricated PAAm disks of thickness 6 mm and diameter 60 mm. The hydrogel disks were fully immersed in deionized water and placed between acrylic plates affixed with sandpaper to ensure a large static friction, as illustrated in Fig.~\ref{fig:normal force vs strain}(a). The PAAm disks were initially compressed with a normal stress of $\approx$ 2-3 kPa, corresponding to a normal strain of 5\%, and then relaxed for 14 hours. Oscillatory strain sweep experiments were conducted over the range $0.006\leq \gamma\leq 0.32$, spending 25 s per point. An oscillatory frequency of 1 Hz was chosen since it is much faster than the diffusive relaxation timescale. The change in normal stress, $\Delta \sigma_N$, is shown in Fig.~\ref{fig:normal force vs strain}(b) for multiple experiments and four different monomer concentrations. All data show remarkable agreement with $\sigma_N \propto \gamma^2$ (Eq.~\ref{normal stress tighe}) over more than a decade in strain. Using Eq.~\ref{normal stress tighe} and accounting for the average dilatant effect over the plate area \cite{supp}, we found that $R_v$ ranged from 13 kPa to 592 kPa. This coefficient can be theoretically described by Eq.~\ref{reynolds coeff. tighe}, which relies on Maxwell relations for the derivatives of elastic moduli in a hyperelastic material \cite{tighe2014shear}.  

A hydrogel is necessarily more complex due to its poroelastic nature, i.e., there are time dependent effects. Since the data from Fig.~\ref{fig:normal force vs strain}(b) starts from a fully-relaxed state of compression, we chose to measure the complex elastic shear modulus, $|G^*| = \sqrt{G'^2 + G''^2}$, and the Young's modulus, $E$, for fully relaxed hydrogels. The moduli of PAAm hydrogels can vary widely depending on cross-linker concentration and environmental conditions \cite{denisin2016tuning}. We choose a wide range of samples (Fig.~\ref{fig:normal force vs strain}(b)) to illustrate the robustness of shear-induced dilation. We found that across multiple samples, the data could be reasonably collapsed by normalizing $\Delta\sigma_N$ by $|G^*|$, which may be expected by the modulus dependency in Eq.~\ref{reynolds coeff. tighe}. The collapse is shown in Fig.~\ref{fig:normal force vs strain}(c). Since $|G^*|$ can decrease with $\gamma$ (Fig.~S2), normalizing the data shifts their vertical position and also leads to a more consistent power law relationship, $\Delta\sigma_N/|G^*|\approx C\gamma^2/4$, where $C\approx23.5$, as shown by the solid line.

Figure~\ref{fig:normal force vs strain}(d-e) shows $|G^*|$ as a function of $\sigma_N$, and $\sigma_N$ as a function of the normal strain for a 12\% PAAm gel. Assuming a linear relationship for both, we extracted $d|G^*|/d\sigma_N=0.93$ and $E=23.0$ kPa from the slopes of each fit, respectively. To account for viscoelasticity, $|G^*|$ was used instead of $G$ \cite{fall2022tuneable}, and the measurement of $E$ accounted for the no-slip boundary conditions on the confining parallel plates (Eq.~S5 \cite{supp, williams2008using}). With these values, Eq.~\ref{reynolds coeff. tighe} predicts $R_v=13.1$ kPa. This should be compared to 14.2 kPa for the 12\% PAAm hydrogel shown in Fig.~\ref{fig:normal force vs strain}(b). The agreement is remarkable given that Eq.~\ref{reynolds coeff. tighe} relies solely on elastic energy \cite{tighe2014shear}, whereas the free energy of a hydrogel contains both elastic and mixing free energies \cite{doi2009gel,flory1943statistical}. 

Lastly, to demonstrate that this increase of $\Delta \sigma_N$ upon applied strain does not lead to an eventual negative normal stress as observed by de Cagny et al. \cite{de2016porosity} in low modulus PAAm hydrogels, we compressed an 8\% PAAm slab for 14 hours, then applied a constant shear stress of 150 Pa for an additional 8 hrs, as shown in Fig.~\ref{fig:normal force vs strain}(f). We observed an exclusively positive normal stress response during this period, although concomitant with the continuing relaxation from compression. This further demonstrates that applied shear stress, regardless of the source, leads to a bulk dilational effect in stiff hydrogels.

\section{Summary and Outlook}
We provided compelling evidence that stiff hydrogels exclusively dilate due to shear in both tribological and rheological experiments. The dilatancy was observed across short and long timescales and can be explained by the bulk, hyperelastic properties of hydrogels. Specifically, the measured Reynolds dilation coefficient ($R_v$) showed excellent agreement with leading order predictions from elasticity theory. Our results contradict prior experiments and theory suggesting purely contractile behavior in network materials and all hydrogels (provided sufficient time has elapsed for poroelastic relaxation). These results may also explain the natural rehydration of stiff tissues such as cartilage during mechanical shearing, where dilation imbibes the surrounding synovial fluid.
Given the contractile behavior observed in low-modulus gels, we expect the Reynolds coefficient $R_v$ to change sign as the polymer network stiffens or softens. This transition should be related to the network micro-structure and cross-linking density. 
%Additionally, the micro-structure can evolve over time and depends on shearing history, giving rise to plasticity observed in experiments of long timescales. 
Although purely elastic models such as random spring networks can capture many mechanical properties of hydrogel materials, they fail to capture shear-induced dilatancy \cite{baumgarten2018normal}. While it is natural to assume that the amount of pre-compression (pre-stress) in hydrogels could affect their hyperelastic behavior, as expected in random spring networks, we did not see strong evidence of this in our experiments. We speculate that in dense, stiff, and highly cross-linked hydrogels, entropic effects encompassing both the crowded polymers and the solvent cannot be captured by a simple harmonic spring network. The relationship between hyperelasticity, poroelasticity, and hydrogel micro-structure provides many open questions for future research.

%Leveraging microstructure-based models such as realistic network model \cite{lei2021recent} may provide further insights on hyperelastic swelling. T

%We speculate that entropic part of the free energy in highly-crosslinked gels {\color{red}cannot be represented by simple elastic springs, and other effects which.  An addition of free energy and entropy may shed light on such behavior.  }

\begin{acknowledgements}
We acknowledge Daniel Sussman for insightful conversations, and Eric Weeks for use of the rheometer. This work was supported by the Gordon and Betty Moore Foundation, grant DOI 10.37807/gbmf12256.
\end{acknowledgements}

\section{Supplemental Materials}

\subsection{Hydrogel preparation}

PAA (polyacrylic acid) and PAAm (polyacrylamide) hydrogels were used as model stiff hydrogels. Commercial desiccated PAA spheres (Deco Beads, JRM Chemical, Inc.) were placed in de-ionized (DI) water and allowed to swell for 24 hours. The DI water was changed twice in the subsequent 48 hours to ensure the removal of free monomers and contaminants. PAAm hydrogels were prepared in two shapes: a sphere for tribology and compression measurements, and a disk for rheological shearing measurements. To make PAAm hydrogel solution, we combined DI water with a 29-1 acrylamide monomer and N,N'-Methylenebis(acrylamide) mixture (8-24 wt\%), ammonium persulfate (0.05 wt\%), and TEMED (0.1 wt\%). All chemicals were purchased from Sigma Aldrich. The mixed gel solution was immediately poured into either a custom silicon spherical mold, or an acrylic disk mold. Polymerization occurred at room temperature for approximately 1 hour. Polymerized hydrogels were then immersed in DI water for 48 hours, and the water was changed every 24 hours to assist with removal of unpolymerized monomers and any contaminants.

\subsection{Fitting Diffusion-Driven Relaxation Curves}

In addition to Eq.~4 in the main text, the following functional forms were used for fitting normal force relaxation curves of hydrogel spheres:
\begin{align}
    \dfrac{F_N(t) - F_\infty}{F_0 - F_\infty} &= 1 - \dfrac{2.56(Dt/a^2)^{\beta}}{1+2.56(Dt/a^2)^{\beta}} \label{fitform2}\\
    \dfrac{F_N(t) - F_\infty}{F_0 - F_\infty} &= e^{-(Dt/a^2)^{\beta}}. \label{fitform3}
\end{align}
We obtained the contact radius $a$ through optical exclusion microscopy, as done in prior work \cite{cuccia2020pore}. To obtain the Young's modulus ($E$), we measured the relationship between the contact radius and the applied normal force, as assumed that Hertzian contact theory is valid \cite{cuccia2020pore}:
\begin{align}
    a^3 = \dfrac{3 F_0 R}{4 E^*}.
    \label{hertz}
\end{align}
Here, $R$ is the hydrogel sphere radius and $E^*$ is the hydrogel's reduced average modulus,
\begin{align}
    E^* =  \dfrac{E}{1 - \nu^2},
    \label{avg modulus}
\end{align}
 where $\nu$ is the Poisson's ratio. Knowing $a$, $F_0$, and $R$, we calculated $E^*$ of a PAA sphere by plotting $F_0$ vs. $a^3$, as shown in Fig.~\ref{fig:hertzian contact}. We obtained $E^* = 42.7$ kPa. Assuming that $\nu=0.4$, the resulting Young's modulus is $E=35.7$ kPa.

\subsection{Interpreting Young's modulus and Reynolds coefficient from rheological measurements}

When compressing a thin disk of elastic material in a parallel-plate rheometer, the ratio between the average normal stress ($\sigma_N$) and the normal strain ($\xi$) is not solely determined by the Young's modulus. If the material is allowed to freely slip along the plates and expand radially, then $d\sigma_N/d\xi=E$. However, if the material is fixed against the plates (i.e. with sandpaper) and cannot slip, then Poisson's ratio and the disk aspect ratio need to be considered. The slope of the normal stress versus normal strain is \cite{williams2008using} 
\begin{align}
    \dfrac{d\sigma_N}{d\xi} = E\dfrac{1 + 3\nu \Bigl(\dfrac{1 - \nu}{1 + \nu}\Bigl)S^2}{1 + 3\nu(1-2\nu)S^2},
    \label{lower bound E}
\end{align}
where $S=A/h$ is the aspect ratio of the gel disk, $A =$ 30 mm is the disk radius, and $h = 6$ mm is the disk thickness. We also measure the shear modulus $G$, which is related to $E$ and $\nu$ from linear elasticity theory:
\begin{align}
    E = 2G(1+\nu).
    \label{upper bound E}
\end{align}
With these two equations, we can solve for 2 unknowns: $E$ and $\nu$. For the 12\% PAAm hydrogel disk shown in Fig.~4C-D in the main text, $E$ = 23.0 kPA and $\nu$ = 0.4.

\begin{figure}
\renewcommand{\thefigure}{S1}
\centering
\includegraphics[width=\columnwidth]{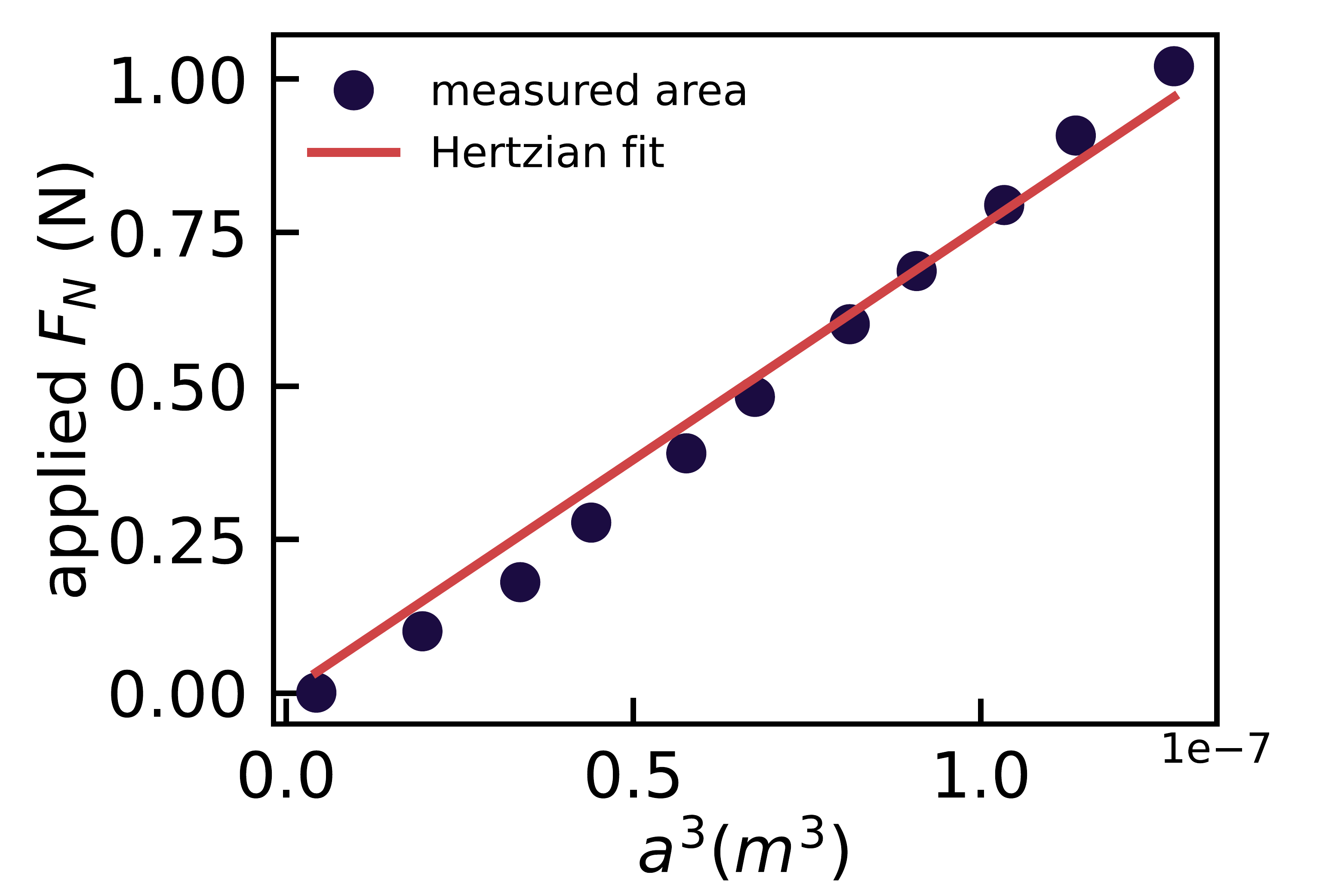}
\caption{Applied normal force versus $a^3$ for a PAA hydrogel sphere pressed onto an acrylic substrate. The data can be reasonably described by Hertzian contact theory, Eq.~\ref{hertz}. The red line represents a linear fit corresponding to $E^*=$ 42.7 kPa, and thus $E =$ 35.7 kPa (Eq.~\ref{avg modulus}). We note that the experiment was performed quickly so that diffusion-driven relaxation of the hydrogel was not important, in contrast to Fig.~1 in the main text.}
\label{fig:hertzian contact}
\end{figure}

\begin{figure}
\renewcommand{\thefigure}{S2}
\centering
\includegraphics[width=\columnwidth]{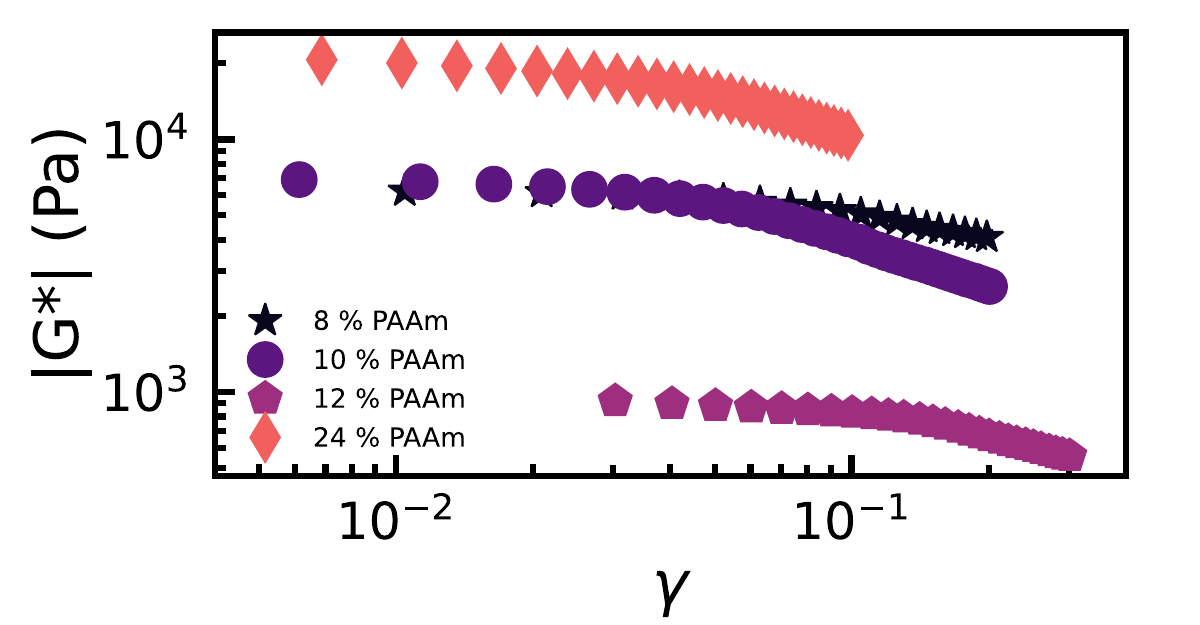}
\caption{Complex modulus $|G^*|$ as measured by oscillatory rheology versus shear strain amplitude $\gamma$ for PAAm hydrogels of various monomer concentrations. The hydrogels were sheared at 1 Hz. Each point is an average of 25 s spent at the corresponding strain value. For all samples, $|G*|$ gradually decreased with strain.}
\label{fig:hertzian contact}
\end{figure}

The rheometer used in our experiments reports the maximum shear strain ($\gamma_{\text{max}}$) at the edge of the parallel plate. In the main text, this is indicated as $\gamma$ for simplicity, i.e., Fig.~4B. However, hyperelastic dilation is a local phenomena, and care must be taken when interpreting rheometer measurements that are averaged over the plate. For example, assuming that the shear strain profile is linear in the radial direction, $\gamma_\text{loc}=\gamma_\text{max}r/A$, where $A$ is the radius of the rheometer plate. Assuming Eq.~2 in the main text applies locally:
\begin{align}
    \Delta \sigma_N = \frac{1}{2}R_v \gamma^2, 
\end{align}
correctly measuring $R_v$ requires integrating over the area of the plate. Specifically:
\begin{align}
    \langle\Delta \sigma_N\rangle = \dfrac{1}{\pi A^2} \int_0^A \dfrac{1}{2}R_v \gamma_\text{loc}^2 2\pi r dr=\dfrac{1}{4}R_v\gamma_\text{max}^2.
    \label{rv correction}
\end{align}
Thus, when measuring $R_v$ from the fits shown in Fig.~4B, we used Eq.~\ref{rv correction}, which includes a factor of 1/4 instead of 1/2.
% Produces the bibliography via BibTeX.

%\begin{figure}
%\renewcommand{\thefigure}{S3}
%\centering
%\includegraphics[width=\columnwidth]{creep.pdf}
%\caption{A series of linearly increasing shear stress periods (1 hr each) resulted in an increase in shear strain and normal force. (a) The applied shear stress as a function of time and the resulted (b) strain $\gamma$ and (c) normal stress difference $\Delta \sigma_N$. Significant creep was observed in both $\gamma$ and $\Delta F_N$ at large stresses. Each symbol represents a repetition of the experiment and the legend indicates when the experiment began with respect to the beginning of the first experiment. The normal strain was held fixed during and between all experiments.}
%\label{fig:hertzian contact}
%\end{figure}

\newpage

\bibliography{ref}

\end{document}